\documentclass{article}

\usepackage[utf8]{inputenc}
\usepackage{titling}
\usepackage{lipsum}
\usepackage[space]{grffile}
\usepackage{graphicx}
\usepackage{url}
\usepackage{amsmath}
\usepackage{amssymb}
\usepackage{hyperref}
\usepackage{float}

\usepackage{verbatim}

\usepackage{xcolor}

\usepackage{pbox}

\newcommand{\RGA}{\mbox{${\mathcal RGA}$}}
\newcommand{\MPRGA}{\mbox{${\mathcal MP}$-${\mathcal RGA}$}}
\newcommand{\UCRGA}{\mbox{${\mathcal UC}$-${\mathcal RGA}$}}

\newcommand{\Tpose}[1]{\mbox{${#1}^{\mbox{\scriptsize{T}}}$}}

\newcommand{\Gm}{\mbox{${\bf G}$}}

\newcommand{\ginv}[1]{{#1}^{\mbox{\tiny -U}}}
\newcommand{\pinv}[1]{{#1}^{\mbox{\tiny -P}}}

\newcommand{\inv}[1]{{#1}^{\mbox{\tiny -1}}}

\title{\bf On Use of the Moore-Penrose Pseudoinverse for\\
Evaluating the RGA of Non-Square Systems}
\author{\Large {Rafal Qasim Al Yousuf} and {Jeffrey Uhlmann} \\
Dept.\ of Electrical Engineering and Computer Science\\
University of Missouri - Columbia}
\date{}

\begin{document}

\maketitle

\begin{abstract}
A recently-derived alternative method for computing the relative gain array (RGA) for singular and/or non-square systems has been proposed that provably guarantees unit invariance. This property is not offered by the conventional method that uses the Moore-Penrose (MP) pseudoinverse. In this paper we note that the absence of the scale-invariance property by the conventional MP-RGA does not necessarily imply a practical disadvantage in real-world applications. In other words, while it is true that performance of a controller should not depend on the choice of units on its input and output variables, this does not {\em necessarily} imply that the resulting MP-RGA measures of component interaction lead to different controller-design input-output pairings. In this paper we consider the application of the MP-RGA to a realistic transfer function relating to a Sakai fractional distillation system. Specifically, for this transfer function we assess whether or not the choice of unit, which in this case relates to temperature, affects the choice of loop pairings implied by the resulting RGA matrix. Our results show that it does, thus confirming that unit-sensitivity of the MP-RGA undermines its rigorous use for MIMO controller design.\\

\begin{footnotesize}
\end{footnotesize}
\end{abstract}

\section{Introduction}

The relative gain array (RGA) has been an important tool for MIMO controller design for over 60 years. Its principal use is to identify strongly-interacting pairs of input and output variables in order to logically decompose that system for the efficient design of an effective controller. The RGA offers a variety of remarkable properties, one of which is invariance with respect to the choice of units on input and output variables \cite{rgajs}. This unit-invariance property is critical in that it guarantees that the design process is not affected by an arbitrary choice of units. Said another way, if the quality of the resulting controller depends on the choice of units, then the only way to have confidence in a given controller design would be to replicate the design process across the entire range of possible units and then assess the relative performance of the resulting designs. If that were necessary then the RGA's role in the process would be entirely superfluous.

As will be discussed, the definition of the RGA involves computation of the inverse of the system plant/gain matrix. This is rarely an issue for systems with matching numbers of input and output variables, i.e., when the gain matrix is square and nonsingular, but in the case of a rectangular gain matrix -- which is necessarily singular -- the inverse does not exist and therefore the RGA cannot be evaluated. A common remedy proposed in the literature is to replace the matrix inverse with the Moore-Penrose (MP) pseudoinverse under an implicit assumption that it represents the best-possible approximation to the conventional RGA when the gain matrix is singular. We refer to this as the MP-RGA because its properties are distinct from those guaranteed by the RGA when it is applicable.  

Although some concerns have been expressed in the literature about the theoretical rigor of the MP-RGA (e.g., in \cite{nrrga}), its continued use was most likely due to the lack of a provably more rigorous alternative. Recently, however, it was formally proven that the MP-RGA does not inherit the unit-invariance property of the RGA, and it was further shown that in order to preserve this property it is necessary to use a different generalized matrix inverse that preserves consistency with respect to diagonal scalings \cite{ucrga}. This generalized inverse is referred to as the UC-inverse \cite{usiam}, which has been successfully applied in a variety of robotic control applications \cite{boz1, boz2}. We refer to its use in computing the RGA as the UC-RGA. The proven unit-invariance property of the UC-RGA, combined with MP-RGA's provable lack of this property, would seem to resolve all questions about how the RGA should be evaluated in the case of a singular gain matrix, but technically it has not been demonstrated that the unit-sensitivity of the MP-RGA actually results in suboptimal pairings when used for controller design. In this paper we formally address this question.

In the next section we discuss the properties of the RGA and demonstrate that the MP-RGA is not scale invariant while the UC-RGA is. We then proceed to our main focus, which is to assess whether or not the choice of units actually affects the input-output pairings implied by the resulting MP-RGA matrix in a practical context. This assessment is performed using a chemical distillation system from the literature that involves differing numbers of inputs and outputs, i.e., has a non-square gain matrix.

\section{The RGA and its Generalizations}

The RGA is a tool that has been used since the late 1960s both for control configuration 
selection in the design of multivariable control systems, and as an interaction measure for
analysis of multivariable control systems \cite{bristol,rgajs}. More specifically,
it provides a relative measure of 
input-output interactions in a multi-input multi-output (MIMO) 
system. For a plant matrix $G$ defining the linear transfer
function from a set of inputs to a set of outputs, the relative gain $\lambda _{ij}$ of the pairing of the $j$th
input $u_{j}$ and the $i$th output $y_{i}$ is defined as the ratio between the loop with all other loops ({\em Open-Loop}) and the same loop with all other loops closed ({\em Closed-Loop}) with perfect control \cite{bristol, Siguard}:

\begin{equation}
	\lambda _{ij}=\frac{\left [ \frac{\partial y_{i}}{\partial u_{j}} \right ]_{\mbox{\em\scriptsize Open-Loop}}}{\left [ \frac{\partial y_{i}}{\partial u_{j}} \right ]_{\mbox{\em\scriptsize Closed-Loop}}}= \frac{\mbox{\em open-loop gain}}{\mbox{\em closed-loop gain}}
	\label{eq:r2}
\end{equation}

The matrix $\lambda _{ij}$ is referred to as the relative gain
array, and it can be defined and evaluated as:
\begin{equation}
   \RGA(G) ~\doteq~ G \circ \Tpose{(G^{-1})}
\end{equation}
where operator $\circ$ is the elementwise (Hadamard) matrix product. Some of the properties of the RGA are \cite{ucrga}:
\begin{enumerate}
\item $\RGA(G) = \RGA(DGE)$ for diagonal matrices $D$ and $E$. (This diagonal-scale invariance is what guarantees unit invariance, i.e., the interaction measures are dimensionless \cite{chen}.)
\item The sum of elements of any row or column is less than or equal to 1.
\item The sum of all the elements is equal to the rank of $G$.
\item A corollary implied by the above two properties is that the RGA of a full-rank $n\times n$ matrix has unit row and column sums. ({\em Proof}: If the sum of elements is $n$, and no row or column can have sum greater than 1, then each row and column sum must be identically 1 in order for the sum of all elements to be $n$.)
\end{enumerate}

One use of RGA measures of interaction is to reduce controller-design complexity by identifying input-output pairs with minimal interaction, i.e., have similar open-loop and closed-loop gains \cite{we,ba,kk,MC}. The ideal case for pairing is $\lambda_{i,j}= 1$. Cases to be avoided are $\lambda_{i,j}<0$ or $\lambda_{i,j}$ much greater than 1, though the tolerable amount greater than 1 depends on the dimensionality $n$ (or rank) of the system. In practice it can be challenging to balance these rules to produce 1-to-1 mapping of pairs\footnote{Satisfying this 1-to-1 mapping constraint is referred to as an {\em assignment} problem or {\em bipartite matching} problem, which can be efficiently solved if the assignment matrix is appropriately transformed \cite{crc4,crc11}. The challenge for use with the RGA (or its generalizations) is transforming the matrix to reflect the priorities of the pairing-selection rules.}. This will be seen in the distillation example of the next section.

In the case of singular $G$, the inverse $G^{-1}$ must be replaced by a generalized matrix inverse. In the literature, this historically has been taken as the Moore-Penrose generalized inverse, sometimes referred to as the {\em pseudoinverse} \cite{changyu}. However, there exist infinite families of generalized matrix inverses, each of sacrifices certain properties of a true matrix inverse \cite{ben,jkusim}. It is likely that the choice to use the MP pseudoinverse of $G$, which we denote with superscript -P as $\pinv{G}$, in the computation of the RGA was due to its widespread familiarity in applications involving squared-error minimization (MMSE). Unfortunately, the MP inverse does not provide consistency with respect to scaling by nonsingular diagonal matrices $D$ and $E$:
\begin{equation}
\pinv{(DGE)} ~ \neq ~ E^{-1}\cdot \pinv{(G)}\cdot D^{-1}.
\end{equation}
which is why the MP-RGA is not unit invariant. The derivation of a diagonal-consistent generalized matrix inverse is relatively recent \cite{usiam}, and exploits the use of canonical scaling (which has also found applications in image processing \cite{aktar}). This unit-consistent (UC) generalized inverse, which we denote with superscript -U, satisfies
\begin{equation}
\ginv{(DGE)} ~ = ~ E^{-1}\cdot \ginv{(G)}\cdot D^{-1}
\end{equation}
and consequently preserves unit-invariance when used in the RGA computation \cite{ucrga}. In other words, the UC-RGA preserves all of the previously enumerated properties of the RGA whereas the MP-RGA fails to satisfy the unit-invariance property.

The MP-RGA, UC-RGA, and RGA are equivalent when $G$ is nonsingular because
$\pinv{G}=\ginv{G}=\inv{G}$. In the $2\times 2$ case, singular $G$ is either 
the zero matrix or has rank $1$, thus constraining the pair interactions
to be equivalent up to a scalar factor. Because of this degeneracy, 
application of the UC-RGA to a $2\times 2$ rank-1 $G$ gives
\begin{equation}
   \UCRGA(\Gm) ~=~ \begin{bmatrix}0.25 & 0.25\\0.25 & 0.25\end{bmatrix}. 
\end{equation}
A more revealing example\footnote{This and the following examples in this section
are based on examples from \cite{ucrga}.} of the difference between the
UC-RGA and the MP-RGA is the case of 
\begin{equation}
   \Gm ~=~ \begin{bmatrix}
                        1 & 1 & 1\\
                        1 & 1 & 1\\
                        1 & 1 & 1
                  \end{bmatrix}, 
\end{equation}
for which all of the interactions among inputs and outputs
are equivalent, and the MP-RGA and UC-RGA solutions
are identical and reflect this: 

\begin{equation}
    \MPRGA(G)~=~\UCRGA~=~
    \frac{1}{9}\cdot
    \begin{bmatrix}
          1 & 1 & 1\\
          1 & 1 & 1\\
          1 & 1 & 1
    \end{bmatrix}
\end{equation}
where, as expected, all of the interaction values are 
the same (and sum to $1$, the rank of $\Gm$).
However, if the first row and column are scaled
by $2$:
\begin{equation}
   G ~=~ \begin{bmatrix}
                        4 & 2 & 2\\
                        2 & 1 & 1\\
                        2 & 1 & 1
                  \end{bmatrix}, 
\end{equation}
the UC-RGA result is unaffected, as should be
expected, whereas the MP-RGA gives
\begin{equation}
  \MPRGA(G)~=~\frac{1}{9}\cdot
                 \begin{bmatrix}
                        4 & 1 & 1\\
                        1 & 1/4 & 1/4\\
                        1 & 1/4 & 1/4
                  \end{bmatrix}, 
\end{equation}
which clearly violates unit-scale invariance. The
more general case of rectangular $G$ can be sanity-checked
using an arbitrary nonsingular $3\times 3$ matrix 
\begin{equation}
   A ~=~ \begin{bmatrix}
                        7 & 4 & 8\\
                        7 & 2 & 5\\
                        3 & 8 & 8
                  \end{bmatrix} 
\end{equation}
and the same matrix after diagonal scaling:
\begin{equation}
   B ~=~ \begin{bmatrix}
                        21 & 16 & 16\\
                        21 & 8 & 10\\
                        9 & 32 & 16
                  \end{bmatrix}
\end{equation}
In other words, $B=DAE$, so their RGAs
are the same:
\begin{equation}
   \RGA(A)=\RGA(B) = \begin{bmatrix}
                        \text{-}2.47 & \text{-}2.41 & ~5.88\\
                        ~3.29 & ~0.94 & \text{-}3.24\\
                        ~0.18 & ~2.47 & \text{-}1.65
                  \end{bmatrix} 
\end{equation}
As shown in \cite{ucrga}, application of the UC-RGA to 
the block-rectangular matrix $[A~B]$ produces
\begin{eqnarray}
\UCRGA([A~B]) & = & [\RGA(A)~\RGA(B)]\\
~ & = & [\RGA(A)~\RGA(A)] \\
~ & = & [\RGA(B)~\RGA(B)]\\
~ & = &   \frac{1}{2}\cdot
      \begin{bmatrix}
         \text{-}2.47 & \text{-}2.41 & ~5.88 & \text{-}2.47 & \text{-}2.41 & ~5.88\\
         ~3.29 & ~0.94 & \text{-}3.24 & ~3.29 & ~0.94 & \text{-}3.24\\
         ~0.18 & ~2.47 & \text{-}1.65 & ~0.18 & ~2.47 & \text{-}1.65
       \end{bmatrix}
\end{eqnarray}
\vspace{6pt}
\noindent The corresponding blocks are not the same for the MP-RGA:\\
\begin{equation}
\MPRGA([A~B]) ~=~
   \frac{1}{2}\cdot
      \begin{bmatrix}
         \text{-}4.47 & \text{-}4.54 & ~9.41 & \text{-}0.49 & \text{-}0.28 & ~2.35\\
         ~5.93 & ~1.77 & \text{-}5.18 & ~0.66 & ~0.11 & \text{-}1.29\\
         ~0.32 & ~4.65 & \text{-}2.64 & ~0.04 & ~0.29 & \text{-}0.66
       \end{bmatrix}
\end{equation}
where the two $3\times 3$ blocks are clearly different
despite the fact that the matrix argument has the form
$A~DA E]$. This clearly shows MP-RGA's 
failure to provide scale invariance.

We now consider an example involving a process transfer function with time values measured in seconds:
\begin{equation}
   \label{tfg}
   G (\mbox{\em seconds}) ~=~ 
     \begin{bmatrix}
         16.8 &  3.2 & 10.5 & 0.6\\
          7.1 &  4.2 &  1.9 & 6.3\\   
          2.5 & 20.0 &  9.4 & 3.4
     \end{bmatrix} 
\end{equation}
The loop pairings (highlighted in boldface) determined from the MP-RGA and UC-RGA for this system are:
\begin{equation}
   \label{mpsecs}
   \MPRGA (\mbox{\em seconds}) ~=~ 
     \begin{bmatrix}
         {\bf 0.6420} & -0.0539 &  0.4387 & -0.0267\\
         0.2588 &  0.0171 & -0.1278 &  {\bf 0.8519}\\
        -0.0521 &  {\bf 0.9052} &  0.1904 & -0.0436
     \end{bmatrix} 
\end{equation}
\begin{equation}
   \label{ucsecs}
   \UCRGA (\mbox{\em seconds}) ~=~ 
     \begin{bmatrix}
         {\bf 0.7394} & -0.0366 &  0.3281 & -0.0308\\
         0.0821 & -0.0803 & -0.0420 & {\bf 1.0402}\\
        -0.0483 &  {\bf 0.9329} &  0.1651 & -0.0496     
     \end{bmatrix} 
\end{equation}
It happens in this case that the two methods agree on the best pairings. However, if the time unit is changed from seconds to minutes, i.e., the first column of $G$ in Eq.(\ref{tfg})is scaled by $1/60$, then the new form of $G$ is:
\begin{equation}
   G (\mbox{\em minutes}) ~=~ 
     \begin{bmatrix}
         0.2800 &  3.2 & 10.5 & 0.6\\
         0.1183 &  4.2 &  1.9 & 6.3\\   
         0.0417 & 20.0 &  9.4 & 3.4
     \end{bmatrix} 
\end{equation}

The dimensionless measures of interaction should not be affected by this change in unit of time, but Eqs.\ (\ref{mpsecs}) and (\ref{mpmins}) show that the MP-RGA -- {\em and its implied pairings} -- are in fact changed. By contrast, Eqs.\ (\ref{ucsecs}) and (\ref{ucmins}) show that the UC-RGA is completely unaffected.
\begin{equation}
   \label{mpmins}
   \MPRGA (\mbox{\em minutes}) ~=~ 
     \begin{bmatrix}
        0.0012 & -0.1678 & {\bf 1.1658} &  0.0008\\
        0.0005 & -0.1253 & -0.0023 &  {\bf 1.1272}\\
       -0.0001 & {\bf 1.2929} & -0.1643 & -0.1285)     
\end{bmatrix} 
\end{equation}
\begin{equation}
   \UCRGA (\mbox{\em minutes}) ~=~ 
     \label{ucmins}
     \begin{bmatrix}
         {\bf 0.7394} & -0.0366 &  0.3281 & -0.0308\\
         0.0821 & -0.0803 & -0.0420 & {\bf 1.0402}\\
        -0.0483 &  {\bf 0.9329} &  0.1651 & -0.0496     
     \end{bmatrix} 
\end{equation}

This example demonstrates that the MP-RGA may not provide reliable information for analysis and design of MIMO controllers when $G$ is rectangular. In the next section we consider a more compelling example.

\section{MP-RGA Application to a Crude-Oil Distillation System}

It is straightforward to show mathematically that the MP-RGA is not unit invariant, and in the previous section we provided a simple example showing that its implied pairings can be sensitive to the choice of units. However, it might be speculated that the MP-RGA may tend to exhibit less fragility when applied in the context of complex real-world systems, which typically have structure imposed by physical constraints of the application. This motivates our consideration in this section of a nontrivial control application from the literature.

Our application of interest involves a Sakai crude-oil distillation system involving a non-square plant \cite{Ex1,Ex1A}, and our goal is to determine whether a change of unit, in this case temperature, affects the input-output pairings implied by the resulting MP-RGA matrix. Based on results from the previous section, we should expect this change of unit to result in changes to the resulting MP-RGA matrix. The critical question, though, is whether the current system will exhibit robustness in the form loop pairings that are invariant with respect to those changes.

\begin{figure}
	\centering
		\includegraphics[scale=0.9]{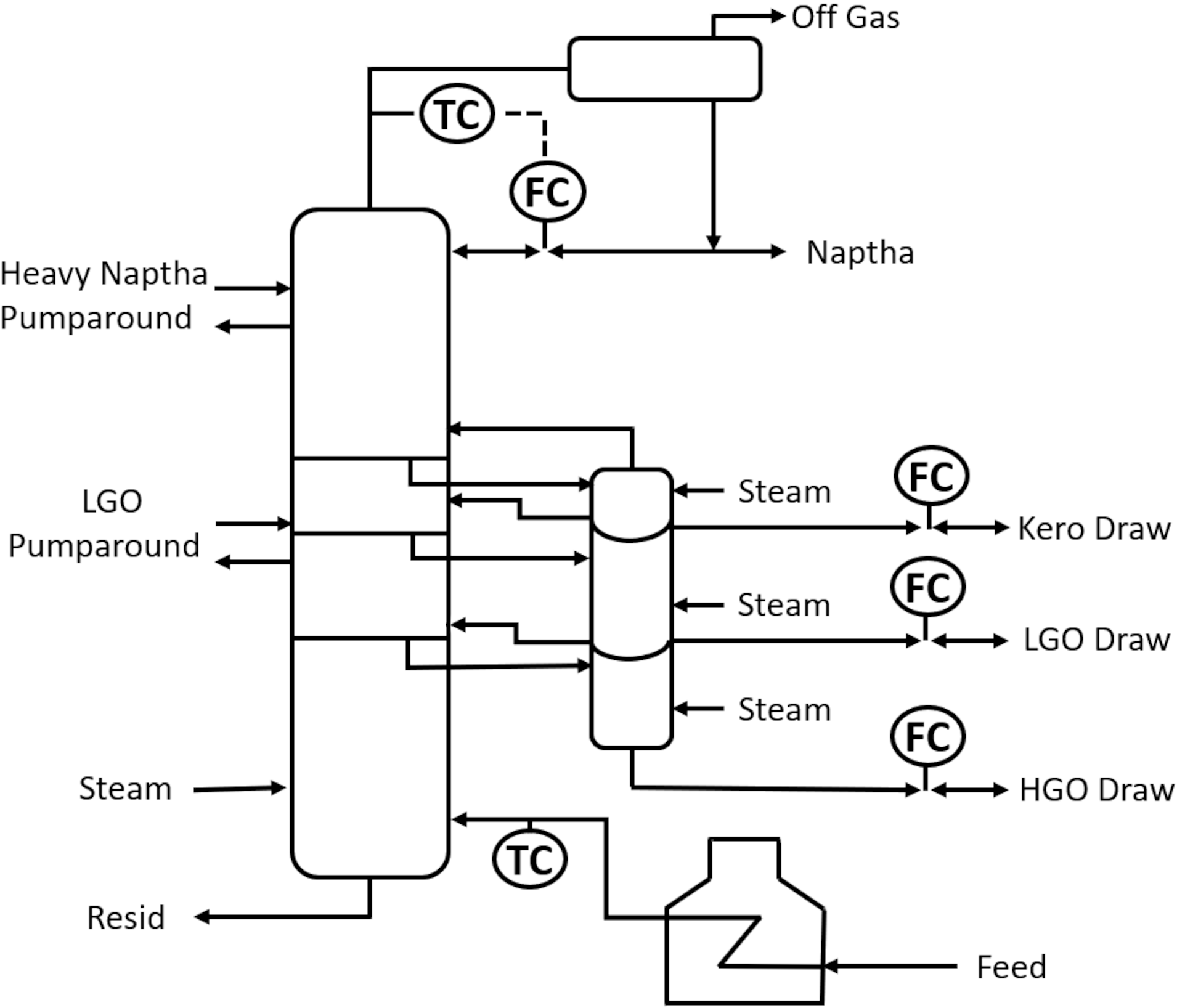}
	\caption{Flow diagram of Sakai's crude distillation unit \cite{Ex1A}.}
	\label{fig:1}
\end{figure}

Figure~\ref{fig:1} is a flow diagram of a Sakai crude distillation unit, which performs preliminary fractional distillation of crude oil prior to the full refining process \cite{Ex1, Ex1A}. 

Manipulated variables are top temperature (u1), kerosene yield (u2), LGO yield (u3), HGO yield (u4) and heater outlet temperature (u5) \cite{Ex1}. Controlled variables are naphtha/kerosene cutpoint (y1), kerosene/LGO cutpoint (y2), LGO/HGO cutpoint (y3), and measured over flash (y4) \cite{Ex1}. A summary of the manipulated variables used by the product quality and the multivariable controllers are given in 
Table\,1 with their nominal operating ranges and their control objectives \cite{Ex1A}.

\begin{table}
\label{sakaivars}
\begin{center}
		\begin{tabular}{lccc}
			\vspace*{5pt}
       {\bf Description of Variables} & {\bf Label} & \pbox{20cm}{\bf Nominal\\ Operating\\ Range} & \pbox{20cm}{\bf Control\\ Objective}\\
      \hline
      ~ & ~ & ~ & ~\\
      Top temperature & $u_1$ & $85-95^{\circ}$C  & --- \\
      Kerosene (kero) yield & $u_2$ & $5-20\%$ & \textemdash \\
      Light gas oil (LGO) yield & $u_3$ & $15-25\%$ & \textemdash \\
      Heavy gas oil (HGO) yield  & $u_4$ & $0-15\%$ & IRV \\
      Heater outlet temperature  & $u_5$ & $340-360^{\circ}$C & IRV \\
      ~ & ~ & ~ & ~\\
      Naptha/kero cutpoint & $y_1$ & $155-185^{\circ}$C & Setpoint \\
      Kero/LGO cutpoint & $y_2$ & $230-260^{\circ}$C & Setpoint \\
      LGO/HGO cutpoint  & $y_3$ & $300-330^{\circ}$C & Setpoint \\
      Measured overflash   & $y_4$ & $1-6\%$C & Zone limits \\
      ~ & ~ & ~ & ~\\
      Kero 5\% point & $r_1$ & $140-170^{\circ}$C & Setpoint \\
      Kero 95\%  & $r_2$ & $220-250^{\circ}$C & Setpoint \\
      LGO cloud point & $r_3$ & $-5-10^{\circ}$C & Setpoint 		
	\end{tabular}  
	\caption{Control variables and control objectives for Crude unit with nominal variable ranges \cite{Ex1A}.}
\end{center}
\end{table}

A summary of the manipulated variables and multivariable controllers are given in Table\,1 with their nominal operating ranges and control objectives \cite{Ex1A}. The process transfer matrix $G(s)$ of the Crude distillation is given in Eq.\ (\ref{eq:r6}). It is a non-square $4 \times 5$ system with 5 inputs and 4 outputs \cite{Ex1,Ex1A}. 

\begin{flushleft}
\begin{equation}
G(s) = \begin{bmatrix}
\frac{3.8(16s+1)}{140s^{2} +14s +1} & \frac{2.9e^{-6s}}{10s+1} & 0  & 0 & \frac{-0.73(-16s+1)e^{-4s}}{150s^{2} +20s +1} \\ 
& & \\
\frac{3.9(4.5s+1)}{96s^{2} +17s +1} & \frac{6.3}{20s+1} & 0  & 0 & \frac{16se^{-2s}}{(5s + 1) (14s + 1)} \\ 
 & & \\
\frac{3.8(0.8s+1)}{23s^{2} +13s+1} & \frac{6.1(12s+1)e^{-s}}{337s^{2}+34s+1} & \frac{3.4e^{-2s}}{6.9s+1} & 0  & \frac{22se^{-2s}}{(5s + 1) (10s + 1)} \\
 & & \\ 
\frac{-1.62(5.3s+1)e^{-s}}{13s^{2}+13s+1} & \frac{-1.53(3.1s+1)}{5.1s^{2}+7.1s+1} & \frac{-1.3(7.6s+1)}{4.7s^{2}+7.1s+1} & \frac{-0.6e^{-s}}{2s+1} & \frac{0.32(-9.1s+1)e^{-s}}{12s^{2}+15s+1} 
\end{bmatrix}
\label{eq:r6}
\end{equation}
\end{flushleft}

Applying the UC-RGA to $G(0)$ with the temperature variable in Celsius gives:
\begin{equation}
\mbox{UC-RGA (Celsius)} = \begin{bmatrix}
{\bf 1.2586} & -0.2889 & 0      & 0    & 0.0303 \\
-0.5381 & {\bf 1.1749} & 0      & 0    & 0.3631 \\
0.4014 & -0.8042 & {\bf 0.8272} & 0    & 0.5755 \\
-0.3561 & 0.4197 & 0.1374 & {\bf 0.7815} & 0.0174
\end{bmatrix} \label{eq:r8}
\end{equation}
where the selected pairings (highlighted in bold) with nonnegative values nearest to 1 align with the first four diagonal elements.  
It must be emphasized that because of its scale-invariance, the UC-RGA result is the same regardless of the unit of temperature. Applying the MP-RGA to $G(0)$ gives two plausible pairing options:
\begin{equation}
\begin{array}{c} \mbox{MP-RGA (Celsius)}\\ \mbox{\small Plausible pairing 1}\end{array} = \begin{bmatrix}
{\bf 1.9147} & -0.9138 &    0      &     0     & -0.0009 \\
-1.1071 & {\bf 2.3221} &    0      &     0     & -0.2150 \\
0.8131 & -1.6290 & 0.6500 &     0     & {\bf 1.1659} \\
-0.7995 & \framebox[1.1\width]{0.9423} & {\bf 0.3086} & 0.5094 & 0.0391 
\end{bmatrix} \label{eq:op1}
\end{equation}
and
\begin{equation}
\begin{array}{c} \mbox{MP-RGA (Celsius)}\\ \mbox{\small Plausible pairing 2}\end{array} = \begin{bmatrix}
{\bf 1.9147} & -0.9138 &    0      &     0     & -0.0009 \\
-1.1071 & {\bf 2.3221} &    0      &     0     & -0.2150 \\
0.8131 & -1.6290 & {\bf 0.6500} &     0     & 1.1659 \\
-0.7995 & \framebox[1.1\width]{0.9423} & 0.3086 & {\bf 0.5094} & 0.0391 
\end{bmatrix} \label{eq:op2}
\end{equation}
where the first two diagonal pairings must be selected to avoid being forced by the 1-to-1 matching constraint to select negative values (or zeros). Numerical confidence in the selections is certainly low because the most confident individual pairing (corresponding to the value 0.9423) is not chosen in either case.  We now consider what happens if the unit of temperature is reduced by a factor of 10, which increases the magnitude of temperature values by a factor of 10:
\begin{equation}
\mbox{MP-RGA (Celsius unit/10)} = 
\begin{bmatrix}
    0.2008  &  {\bf 0.7186}      &   0      &   0  &  0.0806 \\
   -0.1051  &  0.3021      &   0      &   0  & {\bf 0.8030} \\
    0.0411  & -0.0824   & {\bf 0.9823}      &   0  &  0.0589 \\ 
   -0.0404  &  0.0476   & 0.0156  &  {\bf 0.9752}  &  0.0020
\end{bmatrix} \label{eq:cdiv10}
\end{equation} 
This change of unit now results in a numerically-confident set of pairings from the MP-RGA, but it is not consistent with either of the two previous pairings. In summary, the issue here is not which pairing is {\em best}, by whatever measure. What is important is that this example demonstrates unambiguously in a practical application that pairings suggested by the MP-RGA are in fact sensitive to the choice of units.

\section{Discussion}

In this paper we have resolved an open question in the literature regarding the practical implications of the lack of scale invariance provided by the MP-RGA when applied to non-square (or otherwise singular) plant matrix. Specifically, we have shown in a practical example that a simple decimation of the unit of temperature is sufficient to change the set of pairings indicated by the MP-RGA. This tends to strongly undermine confidence in the MP-RGA as a controller design tool when applied with non-square systems. It must be emphasized, however, that our results should not be interpreted as providing general evidence of the practical efficacy of the UC-RGA. Satisfying required mathematical conditions is necessary, but the extent to which that translates into practical relevance is purely speculative \cite{mil}. Therefore, future work is needed to empirically assess the quality of UC-RGA loop pairings across a broader range of control applications.

\section{Future Work}
The RGA has proven to be a useful tool for analysis, structure selection (loop pairings), and controller design for MIMO systems. Most of that utility is due to its various mathematical properties, but there are also interpretational aspects that derive primarily from experience based on its long history of use. For example, the rules for pair selection are far from precise prescriptions, and it has been empirically established that for a given RGA matrix there may be reasons to choose a $\lambda_{ij}$=5 pair over a $\lambda_{ik}$=1 pair because of what selection options would be forced by the latter choice \cite{hovd}. People who have vast experience with the RGA have a strong intuitive feel for this, but the extent to which this intuition holds for applications of the UC-RGA to singular systems is unclear and must be examined.

~\\
\begin{footnotesize}
\noindent {\bf Acknowledgements}: The authors wish to thank Roger Fales and Sigurd Skogestad
for informative conversations relating to the topic of this paper.
\end{footnotesize}

\end{document}